\documentstyle[12pt,epsf]{article}
\let\section=\subsection     \let\subsection=\subsubsection                

\begin{document}
\begin{center}
   {\large \bf INSTANTONS AND MONOPOLES }\\[2mm]
   {\large \bf IN LATTICE QCD }\\[5mm]
   M.~FEURSTEIN, H.~MARKUM and S.~THURNER \\[5mm]
   {\small \it  Institut f\"{u}r Kernphysik, Technische Universit\"{a}t Wien\\
   Wiedner Hauptstra\ss e 8-10, A-1040 Vienna, Austria \\[8mm] }
\end{center}

\begin{abstract}\noindent
   We analyze the interplay of topological objects in four-dimensional QCD
   on the lattice. The distributions of color magnetic monopoles in the maximum
   abelian gauge are computed around instantons in both pure and full QCD.
   We find an enhanced probability for
   monopoles inside the core of an instanton on gauge field average.
   This feature is independent of the topological charge definition used.
   For specific gauge field configurations  we visualize the situation
   graphically.
   Moreover we investigate how monopole loops and instantons are
   correlated with the chiral condensate. 
   Strong evidence is found that clusters of the quark condensate and
   topological objects coexist locally on individual configurations. 
\end{abstract}

\section{Introduction and Theory}

There are two different kinds of topological objects 
which seem to be important candidates for the confinement mechanism:
color magnetic monopoles and instantons.
In lattice calculations we demonstrated that color magnetic monopoles
and instantons are correlated on realistic gauge field configurations 
\cite{wir}. 
Similar phenomena were discussed by other groups 
on semiclassical configurations \cite{andere}.
This might indicate that both confinement mechanisms have the same 
topological origin and that both approaches can be united.
It is believed that instantons and also monopoles can explain chiral
symmetry breaking \cite{SHU88,MIA95}. 
In this contribution we present first results on the local correlation 
of the chiral condensate $\bar \psi\psi (x)$, 
the topological charge density $q(x)$, and the monopole density $\rho(x)$ 
on single gauge fields.

To investigate monopole currents we project $SU(N)$
onto its  abelian degrees of freedom, such that an abelian $U(1)^{N-1} $
theory remains \cite{thooft2}. We employ the
so-called maximum abelian gauge being most favorable for our
purposes.
For the definition of the monopole currents $m_{i}(x,\mu), i = 1,...,N,$ 
we use the standard method \cite{SCH87}. 
From the monopole currents we define the local monopole density as
$ \rho(x) = \frac{1}{ 4 N V_{4}} \sum_{\mu,i} | m_{i}(x,\mu) | \ .  $
%

There exist several definitions of the  topological charge on the lattice.
We use a field theoretic and a geometric charge definition. The 
field theoretic prescriptions are a straightforward  discretization of
the continuum expression.
To get rid of the renormalization 
constants we apply the ``Cabbibo-Marinari cooling method'' 
which smooths the quantum fluctuations of a gauge field. 
The geometric charge definitions interpolate the discrete set of link 
variables to the continuum and then calculate the topological charge 
directly. 
In our studies of the topological charge density $q(x)$ 
we employ the hypercube and plaquette prescription 
for the field theoretic definition \cite{divecchia} and 
the locally gauge invariant L\"uscher charge definition \cite{schier_l}.

To measure correlations between topological quantities
we calculate functions of the type 
\begin{eqnarray} \label{correlations}
\langle q(0) q(d) \rangle \ ,
\langle \rho(0) \rho(d) \rangle \ ,
\langle \rho(0) q^{2}(d) \rangle \ , \nonumber \\
\langle q^{2}(0) \bar \psi \psi (d) \rangle \ ,
\langle \rho(0) \bar \psi \psi (d) \rangle .
\end{eqnarray}
They are normalized after subtracting the corresponding cluster values.

\section{Results}

\begin{figure}[t]
\begin{center}
\begin{tabular}{cc}
\hspace{0.3cm}  Confinement &
\hspace{0.3cm}  Deconfinement\vspace{0.3cm}\\
\epsfxsize=6.0cm\epsffile{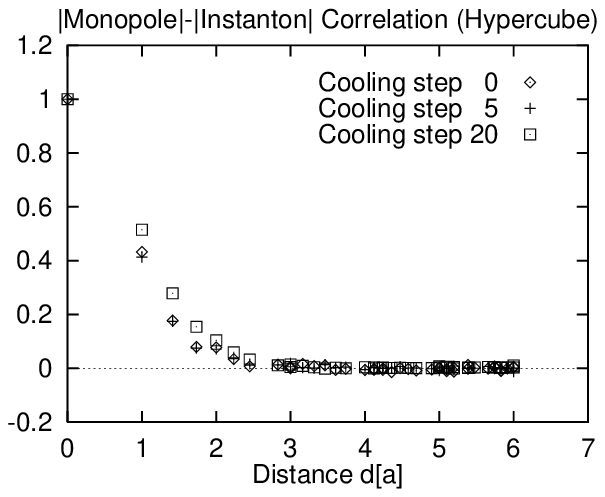}
&
\epsfxsize=6.0cm\epsffile{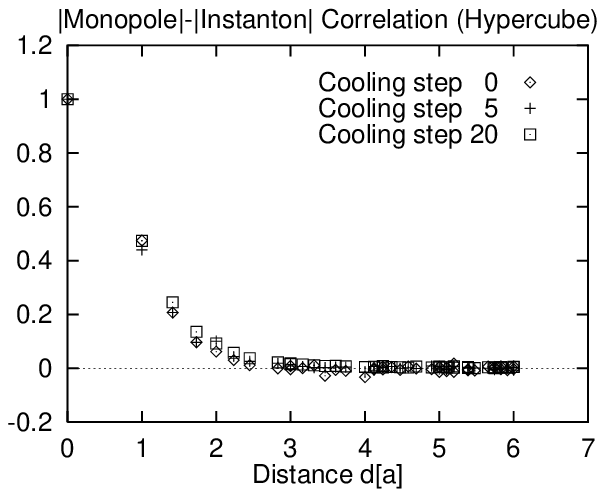} \\
\epsfxsize=6.0cm\epsffile{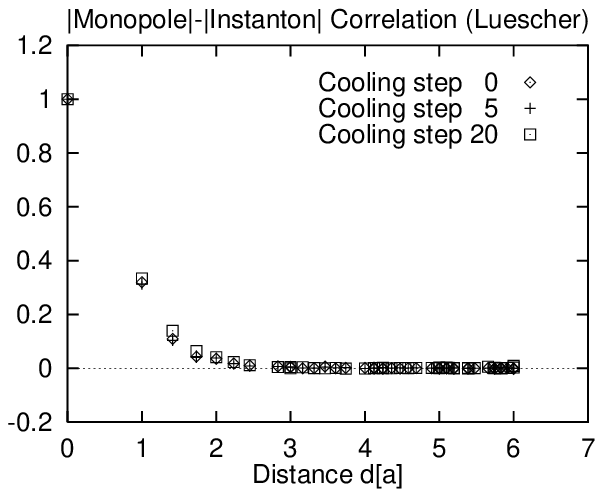}
&
\epsfxsize=6.0cm\epsffile{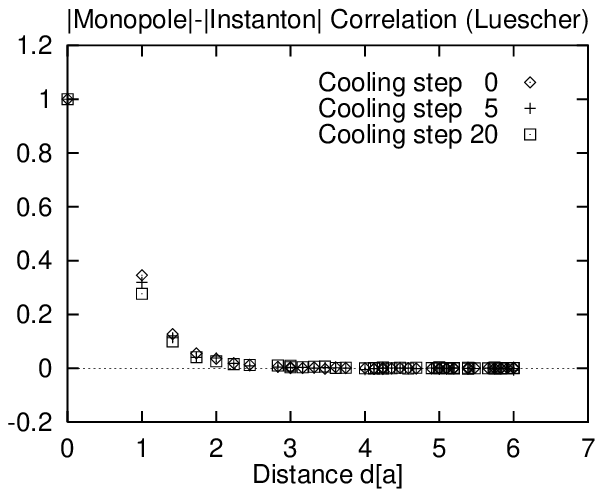} \\ \\
\end{tabular}
\end{center}
\vspace{-0.5cm}
{\baselineskip=12pt
\small Fig.~1.~Correlation functions between the monopole density and the
absolute value of the topological charge density for the hypercube definition
and the L\"uscher definition in the confinement ($\beta = 2.25$)
and the deconfinement phase ($\beta = 2.40$) of pure $SU(2)$ theory.
For both definitions the monopole-instanton correlation functions are almost
invariant under cooling and extend over approximately two lattice spacings.
The correlation functions hardly change across the phase transition. 
\baselineskip=12pt}
\\
\\
{\baselineskip=12pt
\small Table~1.~Screening masses from fits to
exponential decays of the correlation functions between topological objects 
in pure $SU(2)$ theory.
The numbers in brackets are not reliable due to  bad signal-to-noise ratio.
\baselineskip=12pt}
\begin{center}
\begin{tabular}{|lcc|}
\hline
    Confinement &  \hspace{-1.5cm} ($\beta= 2.25$)  &   \\
\hline
 Correlation &  Cool step 5  & Cool step 20  \\
\hline
$q-q$ (L\"u)          & ($1.16 \pm 0.59$) & $1.52 \pm 0.40 $   \\
$q-q$ (hyp)         & ($1.51 \pm 1.12$) & $1.10 \pm 0.48 $   \\
$|q|-\rho$ (L\"u)   & $2.03 \pm 0.22$ & $2.46 \pm 0.19 $   \\
$|q|-\rho$ (hyp)  & $1.86 \pm 0.15$ & $1.94 \pm 0.26 $   \\
\hline
\hline
 Deconfinement &   \hspace{-1.3cm} ($\beta= 2.4$) &  \\
\hline
 Correlation & Cool step 5 & Cool step 20 \\
\hline
$q-q$ (L\"u)          & ($1.94 \pm 0.71$) & $1.00 \pm 0.29 $   \\
$q-q$ (hyp)         & ($2.00 \pm 0.80$) & $1.09 \pm 0.31 $   \\
$|q|-\rho$ (L\"u)   & $1.69 \pm 0.51$ & $1.22 \pm 0.17 $   \\
$|q|-\rho$ (hyp)  & $1.86 \pm 0.32$ & $1.56 \pm 0.20 $   \\
\hline
\end{tabular}
\end{center}
\end{figure}
\noindent 
In Fig.~1 the correlation functions between the monopole density 
and the absolute value  
of the topological charge density for the hypercube and the L\"uscher 
definition are depicted in the confinement (l.h.s.) and the deconfinement 
phase (r.h.s.) of pure $SU(2)$ theory on a $12^{3} \times 4$ lattice with
periodic boundary conditions for several cooling steps. 
Both definitions yield qualitatively the same result.  
For each charge definition
the correlation functions  are almost independent of 
cooling and extend over approximately two lattice units. 
This indicates that 
there exists a nontrivial local correlation between these topological objects
and that the probability for finding monopoles around instantons is clearly 
enhanced. The normalized $\rho |q|$-correlations seem to be hardly 
influenced by the phase transition.  
To gain a more quantitative insight, 
we analyze the correlation functions discussed above by fitting them 
to an exponential function. 
The resultant screening masses in lattice units 
are presented in Table~1 
in both phases for 5 and 20 cooling steps. 
The screening masses computed from the 
correlations between monopoles and instantons turn out to be relatively 
insensitive to cooling and to the phase transition.
The error bars of the masses are large reflecting the large errors 
in the raw data not shown for clarity of plots.

The correlation functions between topological quantities in 
pure $SU(3)$ theory on an $8^{3} \times 4$ lattice
(hypercube definition for $q$)
are shown in Fig.~2 for several cooling steps.
\begin{figure}[t]
\vspace{0.5cm}
\begin{center}
\begin{tabular}{ccc}
\vspace{-0.8cm}
\\
\hspace{-0.8cm} \epsfxsize=4.9cm\epsffile{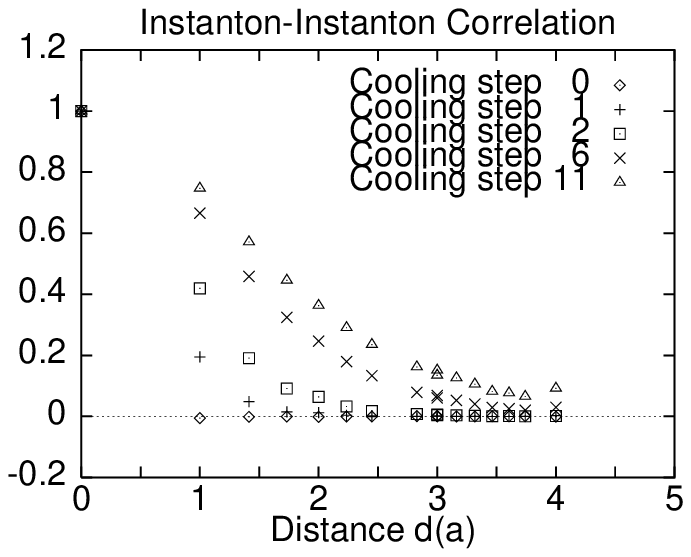}
\vspace{-1.6cm} & 
\hspace{-0.6cm} \epsfxsize=4.9cm\epsffile{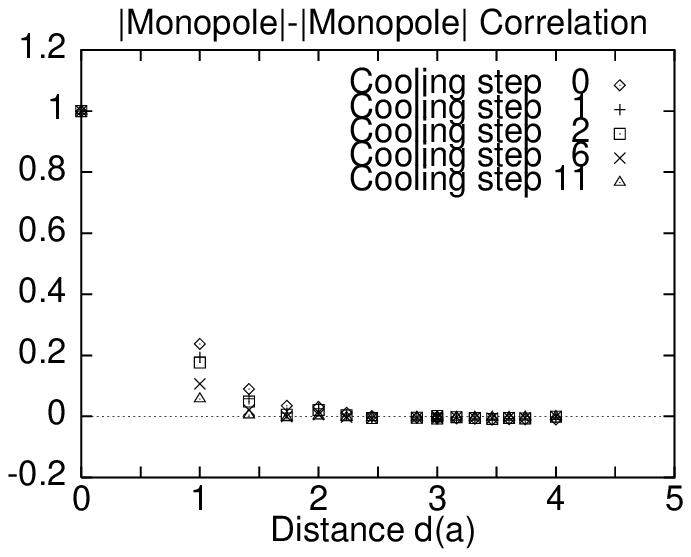}
&
\hspace{-0.6cm} \epsfxsize=4.9cm\epsffile{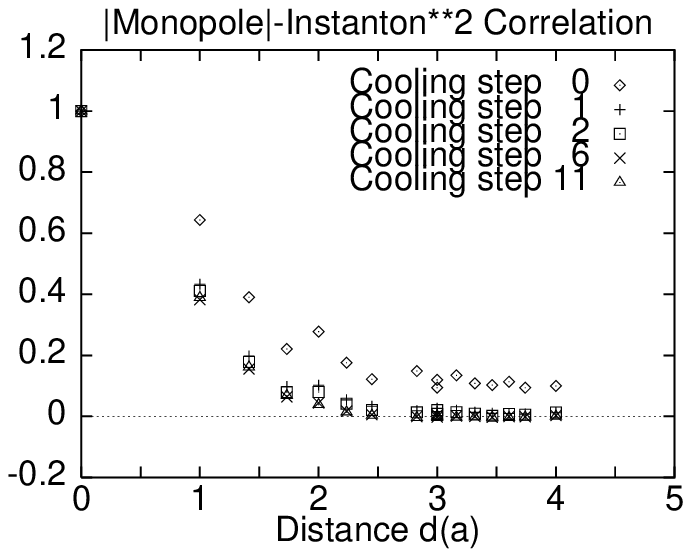} \\ 
{\footnotesize \hspace{2.80cm} (a)} & {\footnotesize \hspace{3.00cm} (b)}
 &{\footnotesize \hspace{3.00cm} (c)}\\
\vspace{0.7cm}
\end{tabular}
\end{center}
{\baselineskip=12pt
\small
Fig.~2.~Correlation functions between topological charge densities
and monopole densities in the confinement phase ($\beta=5.6$) 
of pure $SU(3)$ theory.  
The instanton autocorrelations (a) grow with cooling
reflecting the existence of extended instantons whereas the monopole
autocorrelations (b) decrease since monopoles become diluted.
The correlations between monopoles and instantons (c) show no drastic 
influence from cooling and extend over approximately
two lattice spacings.
\baselineskip=12pt}
\begin{center}
\begin{tabular}{ccc}
\vspace{-0.5cm}
\\
\hspace{-1cm} \epsfxsize=5.2cm\epsffile{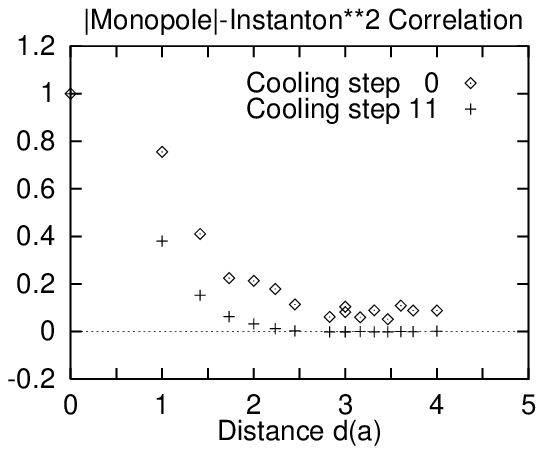}
\vspace{-1.8cm}
&
\hspace{-0.80cm} \epsfxsize=5.2cm\epsffile{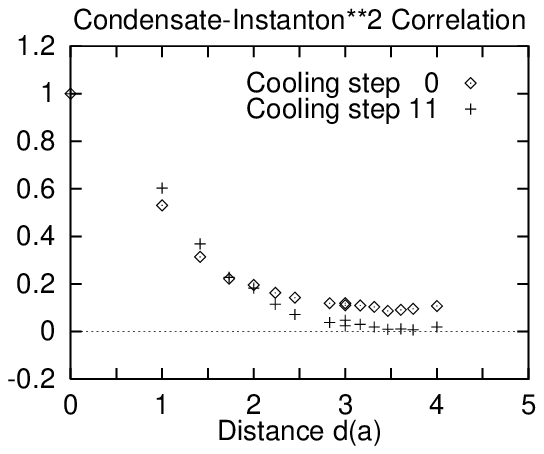}
&
\hspace{-0.90cm} \epsfxsize=5.2cm\epsffile{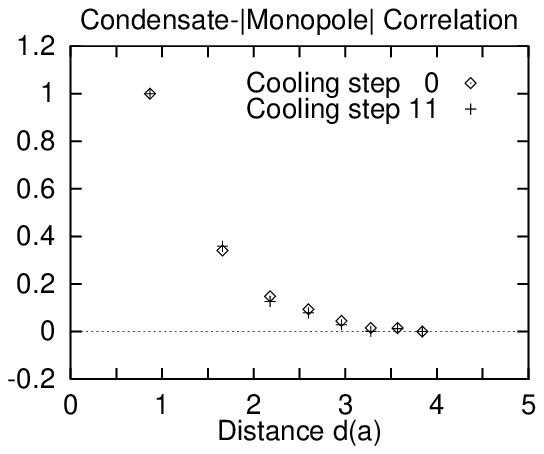}
\\
{\footnotesize \hspace{2.60cm} (a)} & {\footnotesize \hspace{2.80cm} (b)}
 &{\footnotesize \hspace{2.80cm} (c)}\\
\vspace{1.0cm}
\end{tabular}
\end{center}
{\baselineskip=12pt
\small
Fig.~3.~Correlation functions in the presence of dynamical
           quarks in the confinement ($\beta=5.2$). The monopole-instanton
           correlation (a) is similar as in pure $SU(3)$. The
           correlation of the quark condensate and the topological charge (b)
           is cooling dependent, whereas the correlation between the condensate
           and the monopole density (c) is not. All correlations extend
           over two lattice spacings and indicate local
           correlations of the chiral condensate and topological objects.
\baselineskip=12pt}
\end{figure}
The range of the instanton autocorrelation $qq$ (a) being originally
$\delta$-peaked grows rapidly with cooling reflecting the occurance of
extended instantons.
In contrast the $\rho \rho$-correlation (b) decreases
since monopole loops become dilute with cooling. The
$\rho  q^{2}$-correlation (c) seems rather insensitive to cooling and clearly
extends over more than two lattice spacings, indicating some nontrivial
local correlation between monopoles and topological charges.

Figure~3 shows correlation functions of full $SU(3)$ QCD with 3 flavors
of Kogut-Susskind quarks of equal mass $ma=0.1$ in the confinement region.  
The  $\rho q^2$-correlation (a) looks similar to the corresponding function in 
pure QCD (Fig.~2c). The same holds for the instanton and monopole 
autocorrelation functions (not shown). 
In the case of the $\bar \psi \psi q^2$-correlation (b) 
exponential fits show that an increasing number of cooling steps 
results in a narrower  correlation function. 
The $\bar \psi \psi \rho$-correlation (c) on the other hand is not sensitive 
to cooling and has approximately the same exponential decay as the 
$\bar \psi \psi q^2$-correlation after some cooling steps. 

\begin{figure}[h]
\begin{center}
\vspace{0.5cm}
Single instanton solution
\centerline{ \epsfxsize=6.5cm\epsffile{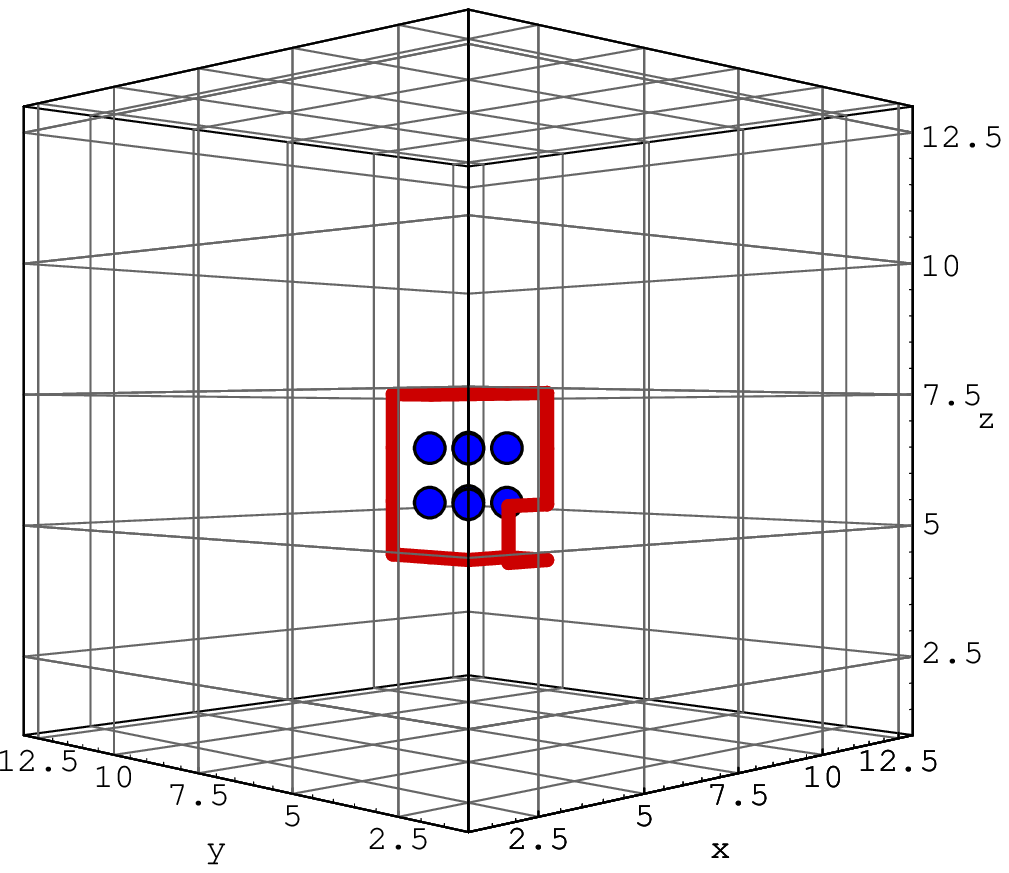} }
\end{center}
{\baselineskip=12pt
\small Fig.~4.~Location of a single instanton (dots) at constant time
imposed on a trivial gauge field configuration.
A closed monopole loop (line) runs around the instanton.
\baselineskip=12pt}
\end{figure}
We now visualize distributions of topological quantities
from individual gauge fields. 
Fig.~4 depicts for a fixed time slice the location of a single 
instanton (dots) which was 
put on a  trivial gauge field configuration artificially. Here a 
purely spatial monopole loop surrounds the instanton (closed line). 

To obtain some insight into the topological correlations, 
Fig.~5 presents a cooling history of an $SU(2)$  
gluon field at a fixed  time slice on a $12^3 \times 4$ lattice.
The topological charge density using the plaquette and the
hypercube definition is displayed for cooling steps  0, 15 and 25.
A dot is plotted if $|q(x)| >0.01$. 
The lines represent the monopole loops.
Without cooling the topological charge distribution cannot be resolved 
from the quantum fluctuations.
Also the monopole loops do not exhibit a  structure.
After 15-20 cooling steps one can identify clusters of topological
charge with instantons. 
At cooling  steps 35-40 the instanton and antiinstanton 
begin to approach each
other until they  annihilate several cooling steps after (not shown).
Monopole loops also thin out with cooling, but they survive in the
presence of instantons. In general, there is 
an enhanced probability that monopole loops exist  in the vicinity of
instantons.
\begin{figure}[t]
\begin{center}
\begin{tabular}{ccc}
\\
{\small Cooling step 0} & {\small Cooling step 15} &{\small Cooling step 25} \\
\hspace{-0.4cm} \epsfxsize=4.6cm\epsffile{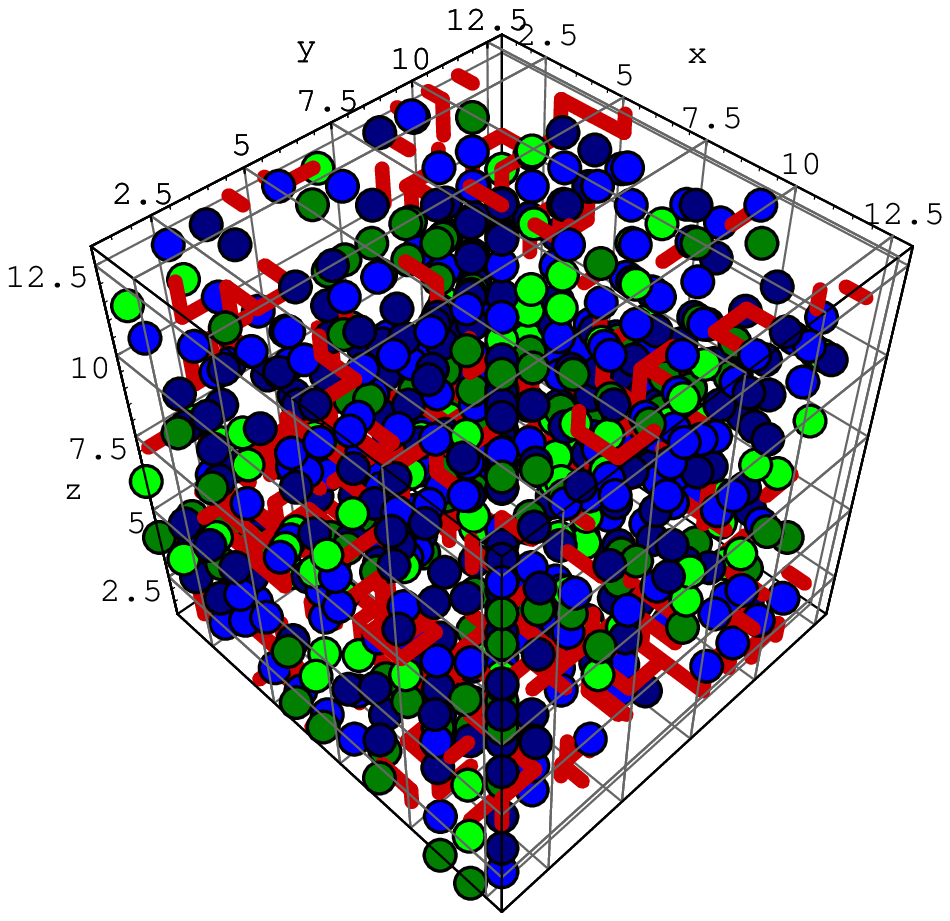}
&
\hspace{-0.3cm} \epsfxsize=4.6cm\epsffile{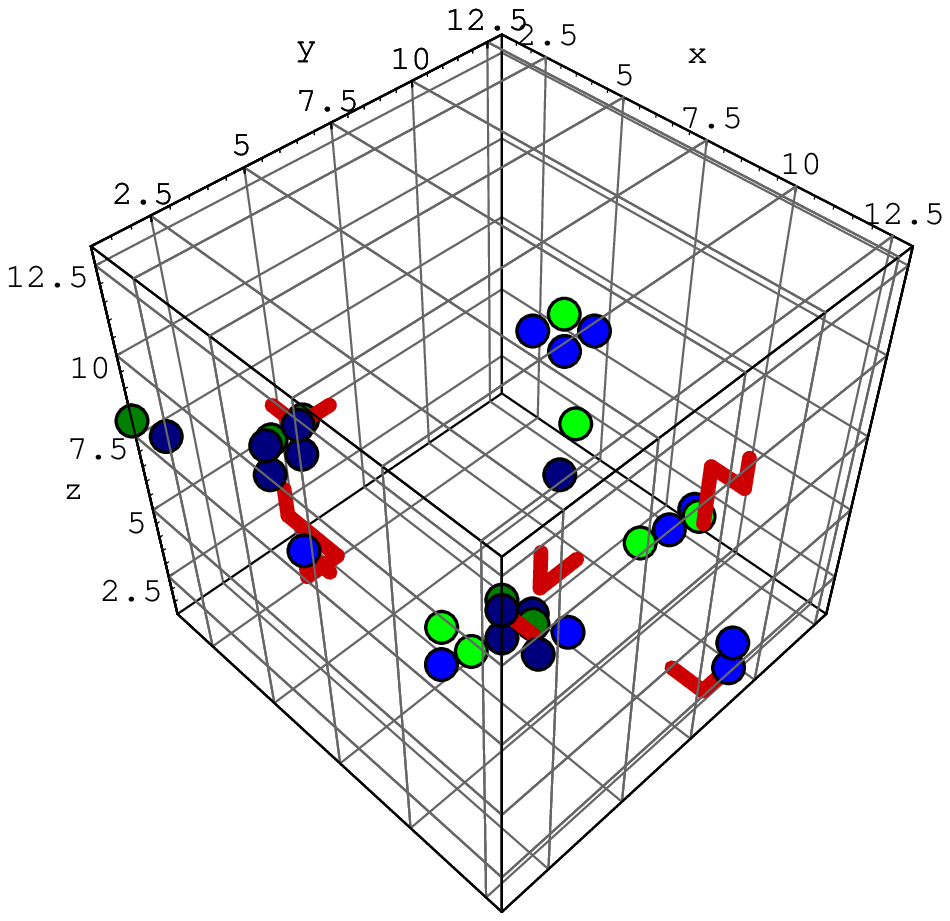}
&
\hspace{-0.3cm} \epsfxsize=4.6cm\epsffile{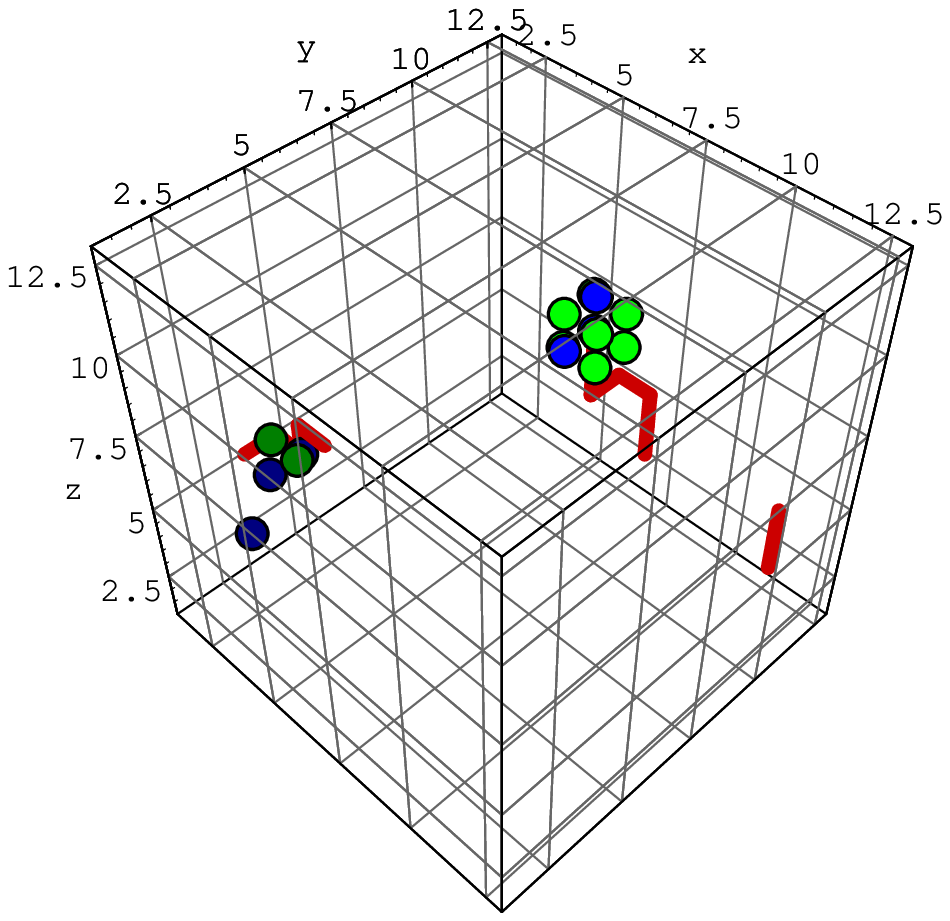}
\\
\end{tabular}
\end{center}
{\baselineskip=16pt
\small Fig.~5.~Cooling history for a time slice of a single gauge field
configuration of pure $SU(2)$ theory. The dots represent the topological 
charge distribution. Monopole loops are represented by lines. 
It can be seen that with cooling instantons evolve from noise 
accompanied by monopole loops in almost all cases. 
Note that black-and-white pictures do not present 
the situation so clearly as color plots. 
\baselineskip=15pt}
\begin{center}
\begin{tabular}{ccc}
\\
{\small Cooling step 1} & {\small Cooling step 5} & {\small Cooling step 10} \\
\hspace{-0.4cm} \epsfxsize=4.6cm\epsffile{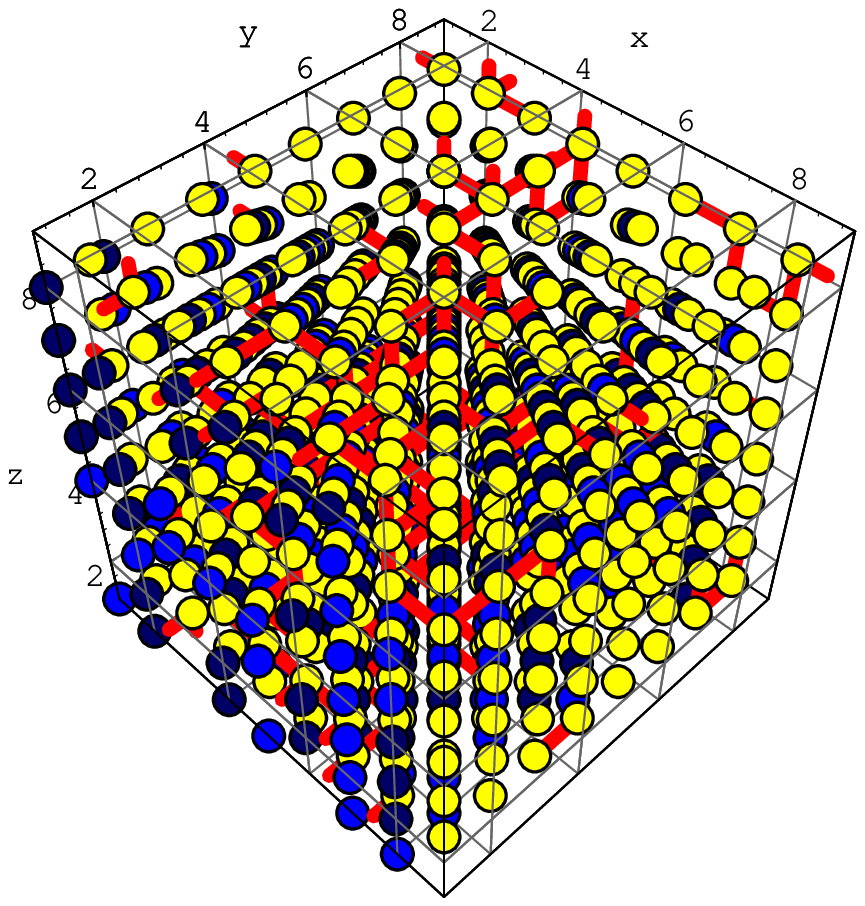}
&
\hspace{-0.3cm} \epsfxsize=4.6cm\epsffile{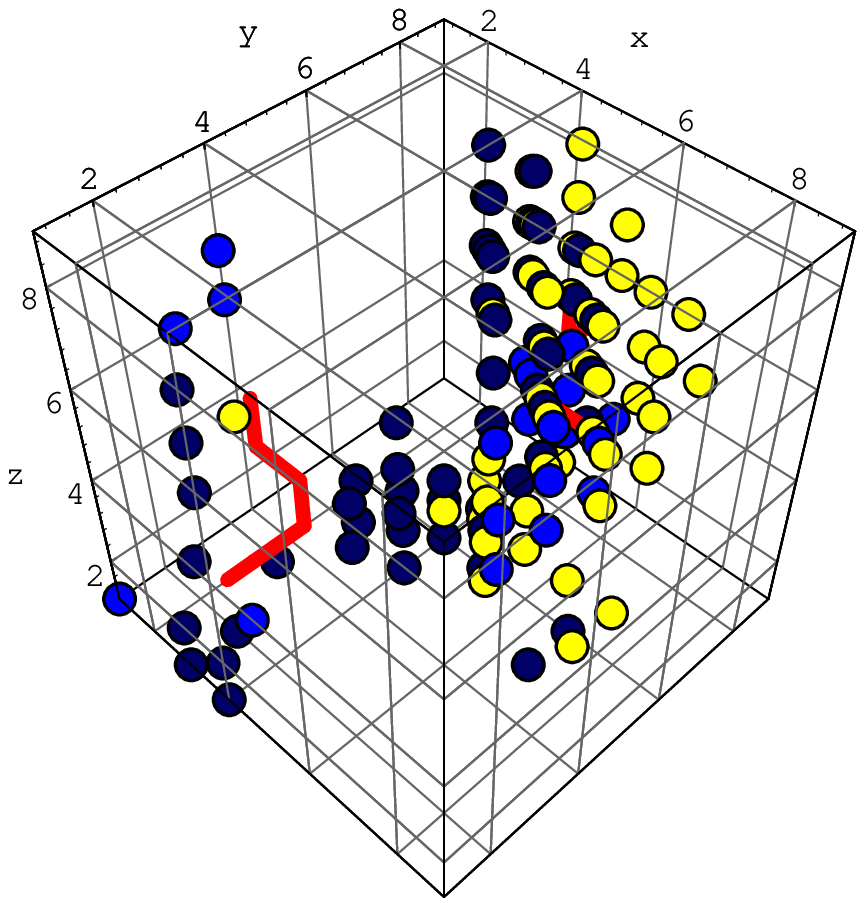}
&
\hspace{-0.3cm} \epsfxsize=4.6cm\epsffile{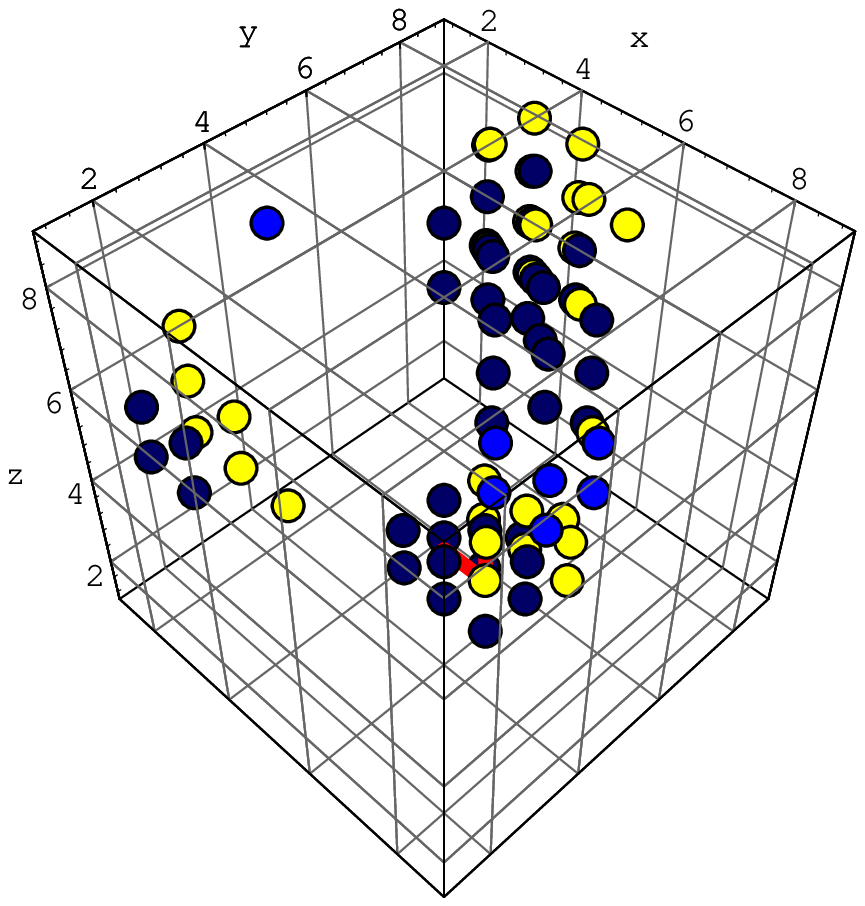}
\\
\end{tabular}
\end{center}
{\baselineskip=16pt
\small Fig.~6.~Cooling history for a time slice of a single gauge field
configuration of $SU(3)$ theory with dynamical quarks. 
The dark dots represent the topological charge distribution and 
the light dots the chiral condensate. Monopole loops are represented by lines. 
It turns out that chiral symmetry breaking occurs at the positions 
of the instantons. 
\baselineskip=15pt}
\end{figure}

We turn to the visualization of distributions of the chiral condensate 
and topological quantities. 
In Fig.~6 a time slice of a typical configuration from $SU(3)$ theory 
with dynamical quarks on an $8^{3} \times 4$ lattice is shown. 
We display the instanton density by dark dots if the absolute value 
$|q(x)| > 0.003$. 
The chiral condensate is indicated by light points if a threshold 
for $\bar \psi\psi (x) > 0.066$ is exceeded. 
Monopoles are represented by lines. 
For clarity of the plots we give only 
one color type of monopole currents, $m_1(x,\mu)$. 
By analyzing dozens of gauge field configurations we found the 
following results. 
The topological charge is covered by quantum fluctuations and 
becomes visible by cooling of the gauge fields. For 
0 cooling steps no structure can be seen in $q(x)$, $\bar \psi\psi (x)$ 
or the monopole currents, which does not mean the absence of correlations 
between them. After 5 cooling steps clusters of 
topological charge and chiral condensate are resolved. 
This particular configuration possesses a positive and a negative 
topological cluster corresponding to an instanton and an antiinstanton, 
respectively. For more than 10 cooling steps both topological charge and 
chiral condensate begin to die out and finally vanish.

\begin{figure}[h]
\centerline{ \epsfxsize=8.0cm\epsffile{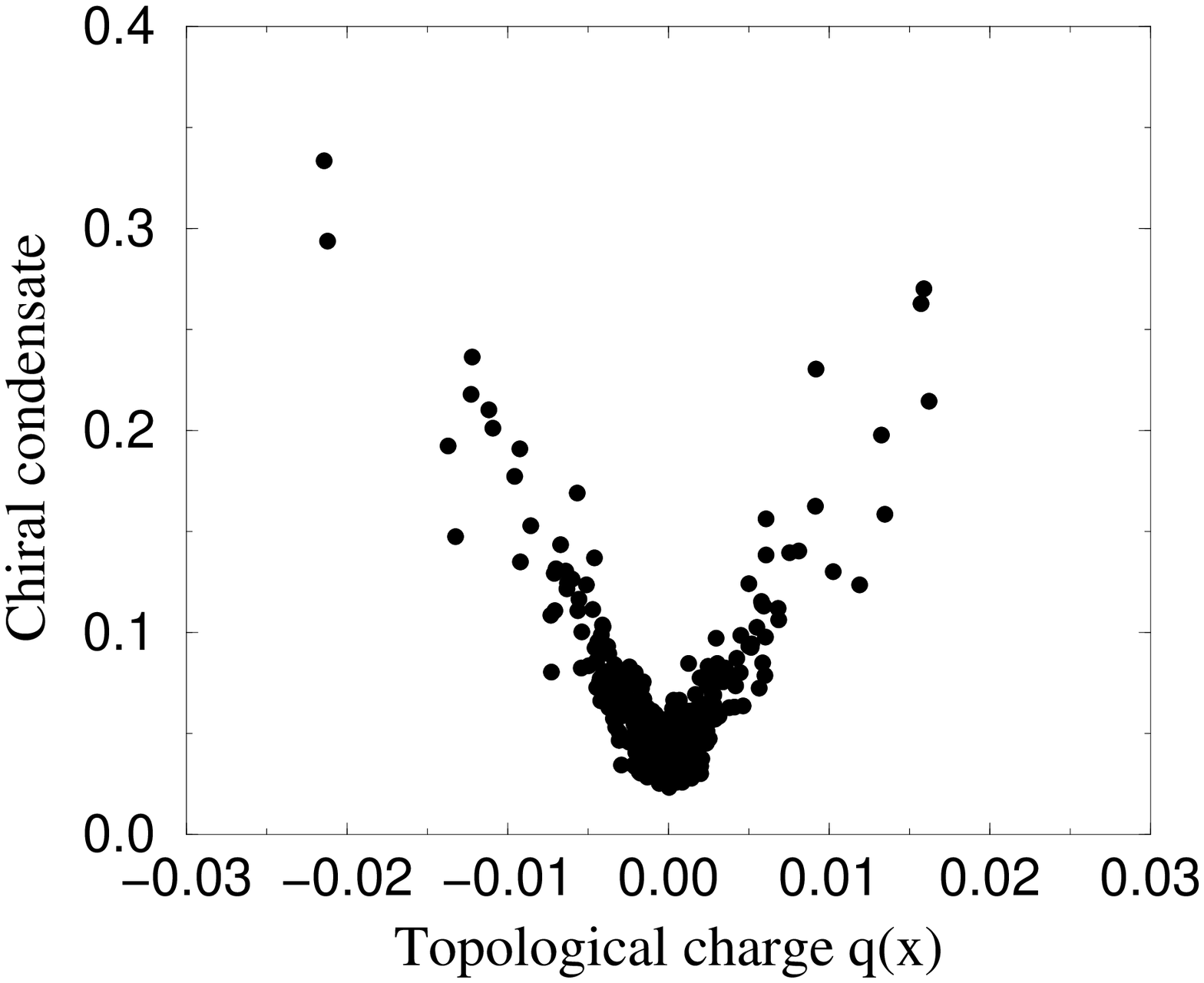} }
{\baselineskip=12pt
\small Fig.~7.~Scatter plot of $\bar \psi\psi (x)$ against $q(x)$ 
in the volume of a single gauge field configuration after 10 cooling 
steps.  
\baselineskip=12pt}
\end{figure}

Figure~6 suggests that the chiral condensate attains its maximum 
values at the same positions as the maximum values of the topological 
charge. This behavior is more clearly shown in Fig.~7 where the 
$\bar \psi\psi (x)$-values are plotted against $q(x)$ for all 
points $x$ in the same configuration at 10 cooling steps. 
Our simulations in the deconfinement phase ($\beta=5.4$) showed 
very little topological activity. In less than one percent 
of the gauge field configurations measured, instantons could be 
identified safely. Also the chiral condensate is considerably lower 
and  does not show the tendency to cluster anymore. 
$\bar \psi\psi (x)$ becomes equally distributed over the lattice 
points. The maximum values of $\bar \psi\psi (x)$ in 
configurations in the deconfinement after 10 cooling steps 
are one order of magnitude smaller. 


\section{Conclusion}
In summary, our calculations of correlation functions 
between topological objects 
and the chiral condensate yield an 
extension of about two lattice spacings. The correlations suggest that the 
chiral condensate takes a nonvanishing value predominantly in the regions  of 
instantons and monopole loops. 
With the visualization of instantons and monopole loops in specific gauge field 
configurations we showed directly that at the sites of instantons 
also monopole loops are present. 
This confirms our conjecture  that monopoles and 
instantons might be two faces of a more subtle fundamental topological 
object, which even might carry an electric charge. 

Further visualization exhibited that the chiral condensate concentrates
around areas with enhanced topological activity (instantons, monopoles).
To our knowledge this observation is the
first direct indication that chiral symmetry breaking occurs
locally in the vicinity  of nontrivial topological structure. 
We found for full $SU(3)$ QCD with dynamical quarks that the clusters 
of nontrivial chiral condensate have a size of about $0.4$ fm, 
which corresponds nicely to the instanton sizes 
observed in the same configurations. In the deconfinement phase 
$\bar \psi\psi (x)$ is significantly lower and looses its cluster
property being present in the confinement regime.

\section*{Acknowledgments}
We thank M.~M\"{u}ller and W.~Sakuler for very helpful discussions. 
This work was partially supported by FWF under Contract No.~P11456-PHY.


\end{document}